\documentclass[journal,twoside]{IEEEtran}

\pdfoutput=1
\usepackage{cite}
\usepackage[pdftex]{graphicx}
\usepackage[cmex10]{amsmath}
\usepackage{algorithmic,algorithm}
\usepackage{array}
\usepackage{mdwmath}
\usepackage{mdwtab}
\usepackage{eqparbox}
\usepackage[tight,footnotesize]{subfigure}
\usepackage{epsfig}
\usepackage{url}
\usepackage{fancyhdr}
\usepackage{mathrsfs}
\usepackage{amssymb}

\makeatletter

\newcommand{\Rmnum}[1]{\expandafter\@slowromancap\romannumeral #1@}
\makeatother

\begin{document}

\title{Internet Resource Pricing Models, Mechanisms, and Methods}

\author{Huan He, Ke Xu, and Ying Liu\\
Institute of Computer Networks\\
Department of Computer Science and Technology, Tsinghua University\\
Email: \{hehuan,xuke,liuying\}@csnet1.cs.tsinghua.edu.cn}

\maketitle

\begin{abstract}
\boldmath
With the fast development of video and voice network applications, CDN (Content Distribution Networks) and P2P (Peer-to-Peer) content distribution technologies have gradually matured. How to effectively use Internet resources thus has attracted more and more attentions. For the study of resource pricing, a whole pricing strategy containing pricing models, mechanisms and methods covers all the related topics. We first introduce three basic Internet resource pricing models through an Internet cost analysis. Then, with the evolution of service types, we introduce several corresponding mechanisms which can ensure pricing implementation and resource allocation. On network resource pricing methods, we discuss the utility optimization in economics, and emphasize two classes of pricing methods (including system optimization and entities' strategic optimizations). Finally, we conclude the paper and forecast the research direction on pricing strategy which is applicable to novel service situation in the near future.
\end{abstract}


\section{Introduction}\label{intro}
\subsection{Background}
\begin{figure}
\centering
\includegraphics[width=0.5\textwidth]{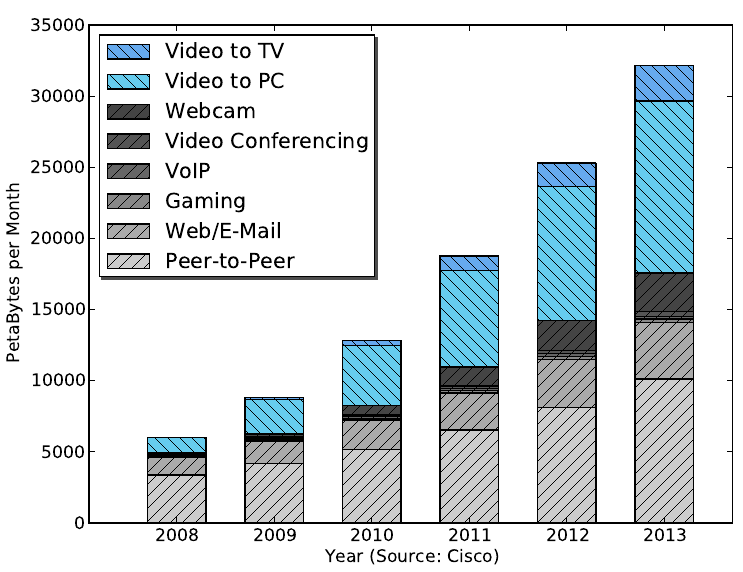}
\caption{Internet video content growth. [7]}\label{fig:1}
\end{figure}

Too many packets will incur network performance degradation, which is called congestion [1]. Congestion is caused by unbalanced resource and traffic distribution, and thus will not be automatically eliminated with the increase of network capacity. In packet switched network, the selfish nature of users makes this happen. As shown by Hardin [2], ``tragedy of commons'' occurs when many individuals share public resources and each holds a selfish objective, which means the loss they bring to others is larger than their own improved benefits. So, if the network is used as public goods, there always exists the possibility that the overall personal excessive usage will cause system performance decline and thus the congestion problem.\\
\indent In recent years, high bandwidth, low latency, low jitter and other higher \mbox{QoS} applications are getting increasingly popular. Thus the surges of network traffic makes network congestion more frequent and serious. Accordingly, the novel content distribution technologies and mechanisms to ensure network QoS are constantly proposed and improved. For the former, commonly, a new layer of network architecture, the application layer network (Overlay Network [3]) is added in the existing Internet to realize the corresponding transmission and QoS control, such as P2P (Peer-to-Peer) [4] and CDN (Content Distribution Networks) [5]. For the latter, mechanisms are developed to work at all levels of QoS control, such as transport layer and network layer concerning network service structures. In short, they both serve network resource management and congestion control.\\
\indent However, on the one hand, network traffic surges and keeps increasing. As Valancius [7] shown in Fig.~\ref{fig:1}, videos and P2P traffic occupy a large part of network resources and will become even more in the coming years. On the other hand, different application layer networks have their own selfish traffic demands and QoS control mechanisms. This makes network management and maintainence increasingly difficult [6]. As an earlier \mbox{best-effort} network service type, Internet Service Providers (ISPs) often meet the increasingly high QoS requirements by upgrading network infrastructure or increasing network capacity. However, in the long run, \mbox{short-term} investments usually bring high cost and fail to satisfy the fast-growing network resource requirement, which is against the healthy network development. Therefore, QoS control technologies need to be introduced in \mbox{best-effort} network. From the perspective of improving network resource usage and management, network designers and ISPs usually passively conduct QoS control based on the existing network traffic, such as congestion control [8]-[10], and traffic engineering [11]. But these often complicate network protocol design and implementation. Proactively setting QoS levels of flows for simple QoS control (priority-based QoS mechanism [13][14]) and designing network architecture to ensure QoS (such as IntServ [15] and DiffServ [16]) have also been studied extensively. But due to some technological limitations and lack of incentives, they have not been implemented throughout the network.

\indent In fact, for network designers, it is very effective to improve network performance using the enhanced transport layer protocol design and related underlayer techniques [9][10]. However, they do not care about the types of high-level applications. Thus the corresponding QoS differentiation is hard to ensure. As a result, promoting reasonable and efficient usage of network resources based on applications is more and more emphasized. And the related service types that can provide different QoS levels on different applications are also under in-depth study. Earlier, priority-based network service layering [13][14] tries to achieve a certain level of packet transmission QoS differentiation based on distinguishing the high-level application characteristics of packets. Then, the proposed IntServ architecture [15] guarantees applications' QoS by per-flow resource reservation, and DiffServ [16] modifies the IntServ architecture using priorities based on aggregated flow control. Theoretically, they can improve network resource-use efficiency, indicating a QoS guaranteed service era is coming. However, in addition to technical difficulty and deployment complexity, they are generally achieving high-priority service QoS guarantee at the expense of low-priority services without congestion elimination attempts in nature. Furthermore, due to the distributed management features of the Internet, ISPs lack adequate enthusiasm to collaboratively improve network performance and efficiency without appropriate incentives. Thus QoS guarantee is difficult to implement in the whole network.
\subsection{Resource Pricing}
From the above discussion, we note that design incentives at economical level to encourage ISPs in improving network performance and directing users to use the resources rationally, will be of great significance in effective network resource management and distribution [39]. Such methods are based on the utility optimization theory in economics, which affects users' demand and belongs to active resource management mechanism. Simply speaking, ISPs can effectively influence users' demands and network resource usage by choosing rational pricing strategies, thus prompting efficient network usage and ensuring network performance. Particularly, as an important auxiliary aspect of technological progress (economic incentives [22]), pricing mechanism studies suited to service type development are also important. Therefore, a complete picture of network pricing should include three aspects: basic pricing models, mechanisms to ensure pricing implementation, and methods determining optimal pricing levels.\\
\indent Specifically, first of all, pricing models decide which factors to charge, or how to evaluate network operating and maintaining costs. Mason and Varian [18][19] analyzed the major fee component from users' cost point of view. This includes: a fixed fee to provide basic service structure costs such as leased lines, routing equipment maintenance, and human resource utilities; marginal costs of access; network expansion costs; marginal costs of sending data packets into the congested network; and social costs that cause negative impact on other users. The authors believe a good price should reflect these costs. So, we introduce three basic pricing models concerning these costs: flat pricing [18], usage pricing[21][22][25] and congestion pricing [18][29]-[37].\\
\indent As applications are simple and resources are sufficient at the beginning of the Internet, it is convenient to charge users using a static flat pricing model, where users have the same usage-irrelative fixed fees with equal access rates. The advantages are that complex audit and statistics are unnecessary, and thus facilitates network users. So, they increasingly enrich network contents. However, too many contents eventually causes network resources lacking.And the defects of flat pricing gradually emerge. For the system, due to lacking of incentives for efficient network resource usage [20] (a lot of bandwidth are wasted by non-critical applications), the overall network performance degrades. For users, the experience deteriorates and the fairness cannot be guaranteed. Obviously, flat pricing is no longer applicable. Thus, a more effective resource pricing model ``usage-based pricing'' was proposed [21]. It pointed out that if the charge is related with usage, fair and efficient use of resources will be promoted to some extent. However, with a further increase in network traffic, the aggravated congestion makes the related pricing a hot research area, resulting in a relatively dynamic pricing model ``congestion pricing'' [18][19] which are studied extensively. Besides, these three pricing models can be used in any combination since they reflect different cost components.\\
\indent As for pricing mechanisms, they mainly aim to address the matching problem between network service types and pricing models. Namely, for different types of network services, we need to select and design suitable pricing models. Good pricing mechanism can set rational price structures for users and ensure pricing implementation with an acceptable technical complexity measure [12]. Generally, in \mbox{best-effort} network, ISPs always adjust the basic pricing model to promote the rational use of resources based on their network capacity, where no additional QoS control mechanisms are conducted. Odlyzko's PMP (Paris Metro Pricing [55]) pricing aims to achieve QoS differentiation and thus enhances efficiency through dividing network into several subnets in \mbox{best-effort} network. However, with the increasing emphasis on applications' QoS and network resource usage efficiency, network designers and ISPs both tend to serve different data streams with different QoS and price levels. Simple priority-based pricing was first proposed by Cocchi et al. [13] [14]. The authors suggested to implement prioritized service using priority field in IP packets, and thus they can conduct service layering and corresponding pricing. Similar thoughts can be found in [42]. With progressive development of various network service types, QoS guaranteed network architectures (such as IntServ and DiffServ) are gradually studied in recent years, followed by corresponding pricing models. QoS based network resource pricing mechanisms are thus formulated [43]-[51][56]-[60]. We discuss pricing models suitable to various service types in Section 3.\\
\indent For the last aspect of pricing strategy, pricing methods applicable to pricing model/mechanism are still an important research aspect. It mainly determines how to set a reasonable price level. An ideal pricing method should be able to set price levels that can control resource usages so as to achieve its pricing objectives while achieving efficiency. Determining prices is usually based on relevant fields of pricing and utility optimization in economic theories under specific market environments. Such work is often based on different market structures (such as monopoly and competitive network) and network service mechanisms (such as \mbox{best-effort} and QoS guaranteed service network). After studying each entity's utility, different theory models are used to describe their interactive optimization processes. The theoretical bases are mainly optimization theory and game theory. Thus there are two major research lines: (1) Studying pricing based on system optimization (Network Utility Maximization, NUM [30][31]) always lies in optimization theory [74]; (2) Studying pricing based on strategic optimizations of ISPs and users. That is, when analyzing each player's decision making, one should take into account effects from strategic behaviors of other players. This work is mainly based on two major theoretical branches of game theory: non-cooperative game theory [75][84] (related models such as in [77]-[82]), and cooperative game theory [83]-[85] (related models such as in [87][91][92]).

\subsection{Organizations}
\begin{figure}
\centering
\includegraphics[width=0.5\textwidth]{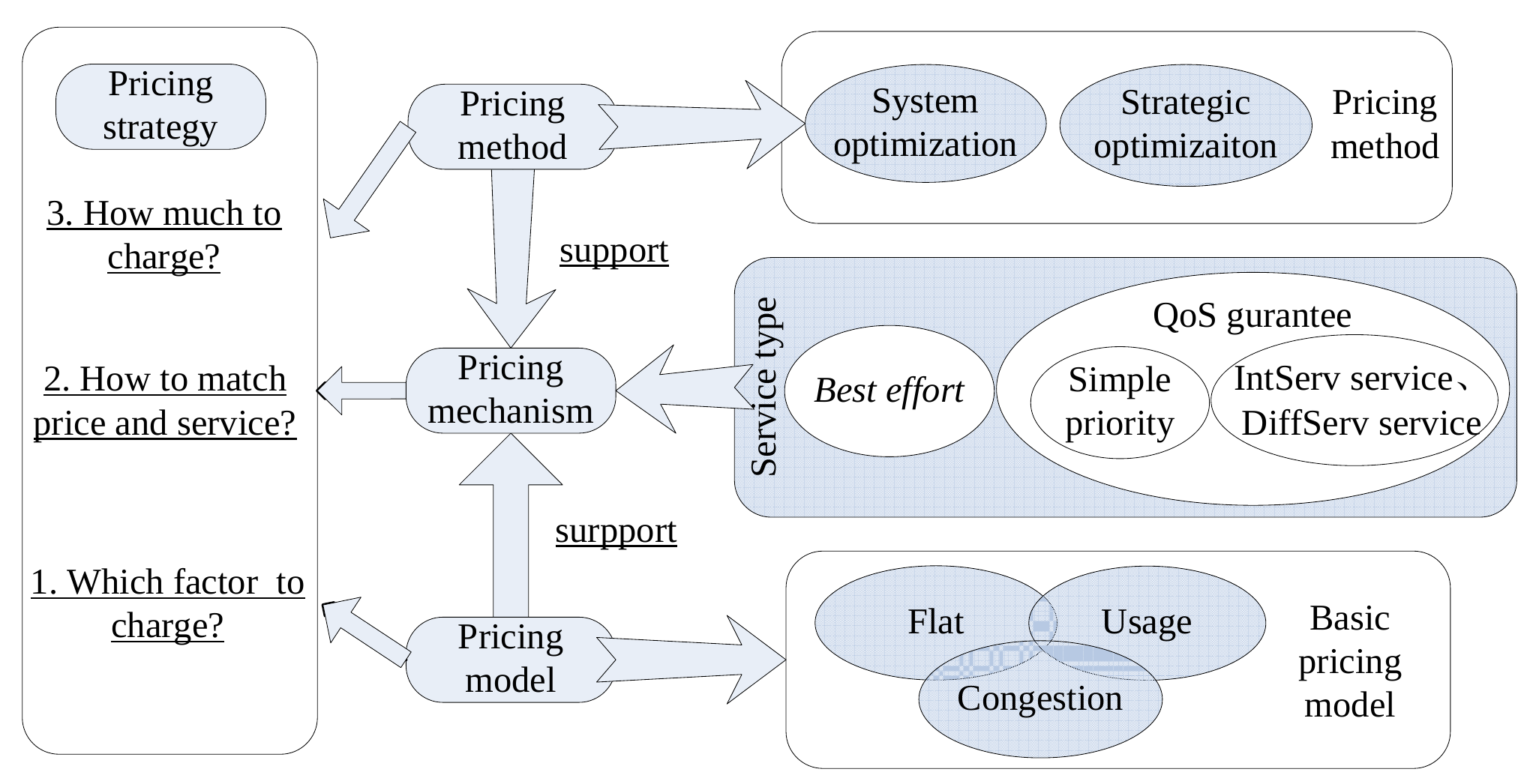}
\caption{The structure of pricing strategies.}\label{fig:2}
\end{figure}

As shown in Fig.\ref{fig:2}, the remainder of the paper presents a detailed survey on Internet pricing development. In Section 2, we present three main pricing models proposed in earlier years. Then, integrated with pricing models, we introduce pricing mechanisms based on two types of services in Section 3. In Section 4, we introduce price level setting methods based on two classes of optimizations, including system optimization and entities' strategic optimizations in different network marketing environments, which can economically incentivize technology development. We classify and compare typical pricing strategies in Section 5 based on different pricing models, serving mechanisms and pricing methods involved. Finally, in Section 6, we conclude the paper, predict the reasonable pricing strategies for new applications and network services, and point out several future research topics.

\section{Basic pricing models} \label{sec:basic}
In the study of pricing models, the main idea is to decide pricing factors based on ISPs' costs. In traditional \mbox{best-effort} network, three basic models can be used for network pricing based on cost analysis. The three models are also important factors in the pricing of subsequent QoS guaranteed network services. This section will inform the three basic pricing models which are gradually evolved in early Internet.
\subsection{Flat pricing} \label{flat}
\begin{figure}
  \centering
  \subfigure[]{
    \label{fig:3} %% label for first subfigure
    \includegraphics[width=0.3\textwidth]{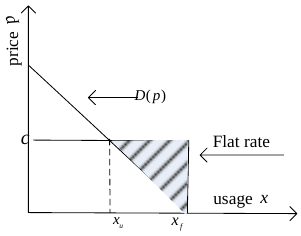}}
  \hspace{0.5in}
  \subfigure[]{
    \label{fig:4} %% label for second subfigure
    \includegraphics[width=0.3\textwidth]{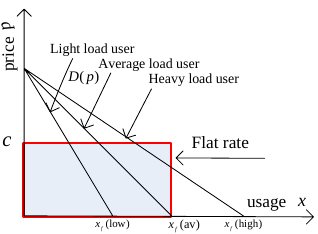}}
  \label{fig:subfig} %% label for entire figure
   \caption{(a) A customer will consume $D(p)=x_u$ unit sat a unit price of $p$, and
  $x_u$ under a flat-rate charge. The shaded area represents the waste. [20] (b) At a unit cost of $c$, the flat-rate charge is the rectangle. The small
  triangle is the value to the light user, and the large triangle is the value to the heavy user. [20]}
\end{figure}

At the early stages of Internet, users use a small quantity of network resources. Thus ISPs aim to attract a large number of users and occupy the market. They generally adopt unified price (or flat fee [18]) to charge users based on access costs, which means in a certain period of time, the users with the same access speed will be charged at the same price. This is especially common in broadband access market.\\
\indent The advantages are as follows. For ISPs, flat pricing is popular, since it is easy to implement and there is no need for complex statistical systems. And for users, the charges can be predicted. However, the more usage, the more obvious drawbacks. On the one hand, due to lack of effective interactions between users and ISPs, users have no incentives or ideas about adapting their usage patterns, making network resources over requested or used. On the other hand, ISPs do not count individuals' resource consumptions and treat equally to users with the same access rate level. This means the overall cost is equally shared by users with different consumptions and thus fairness is hard to guarantee. Meanwhile, as there is no difference in charging users, ISPs lack impetus for upgrading infrastructure or improving QoS, which is not conductive to the progress of network technology and makes system performance degrade.\\
\indent As to the fairness, Edell and Varaiya studied users' reactions on flat pricing through Internet Demand Experiment project (INDEX [20]). They concluded that light-load users compensate the heavy-load ones under flat pricing, which will cause resource waste too. The authors assumed unit usage cost is charged by $c$, and users request $D(c)$ unit resource according to demand curve. As using flat pricing model, the marginal usage cost for users is 0, which makes the demand changed from $D(c)$ to $D(0)$. Estimated by users' practical utilities, the usage over $D(c)$ will cause ${\int}_0^c[D(c)-D(0)]dp$ value loss to users, as the shade shown in Fig.~\ref{fig:3} [20]. In addition, if the flat fee $C$ is charged based on average usage amount, then $C=c\times x_f(av)=c\times D(0)$. All users' payments are shown as the rectangle area in Fig.~\ref{fig:4} [20]. Clearly, the light-load users' payment is more than their gain, while the heavy-load users are on the contrary. This indicates that the former compensates the latter when they share resource costs on average.\\
\indent As discussed above, in \mbox{best-effort} network without additional QoS mechanism, flat pricing model is unable to achieve optimized resource allocation alone. And due to fewer ISPs, the marketization is not obvious, which worsen the situation that ISPs lack incentives to improve network performance. With the development of network applications and the increasingly complex Internet marketing environment, the model will no longer apply. But as one of the referential pricing factors, access charge can be used as a basic guarantee for recovering the fixed costs.

%\begin{figure}
%\includegraphics[width=0.22\textwidth]{fig/fig3.pdf}
%\caption{A customer will consume $D(p)=x_u$ unit sat a unit price of $p$, and $x_u$ under a flat-rate charge. The shaded
%area represents the waste. [20]}\label{fig:3}
%\end{figure}

%\begin{figure}
%\includegraphics[width=0.22\textwidth]{fig/fig4.pdf}
%\caption{ At a unit cost of $c$, the flat-rate charge is the rectangle. The small triangle is the value to the light user,
%and the large triangle is the value to the heavy user. [20]}\label{fig:4}
%\end{figure}

\subsection{Usage pricing}\label{usage}
As the usage and fixed costs have been distinguished and studied separately, usage-based pricing models come into being.
Currence et al. [22] thought usage-based pricing can reflect actual use of network resources and is derived from traditional flat pricing. Simple usage-based pricing uses the amount of upload and download traffic to charge.\\
\indent In practice, China Education and Research NETwork (CERNET) uses full-rate accounting charges for international traffic [24]. In addition to such direct traffic statistics, ISPs in general can use statistical sampling methods to estimate usage, such as the $95^{th}$ percentile pricing which is used as an industry standard. This is in accordance with usage-based pricing, and the peak flow within 5\% of the time (36 hours per month) is free of charge. Many ISP, such as MCI WorldCom and Level (3) Communications, have such peak flow rate based charging standards [22].\\
\indent Usage-based pricing is analyzed and studied by a lot of researchers at early stages of the Internet [12][21]-[25]. The common point is that in general they used supply-demand balance models in economics to describe the interactions between users and ISPs. Edell and Varaiya [20] showed in their experiments that users are highly sensitive to pricing models and price levels. Usage-based charging, can not only enhance usage efficiency of network resource, but also play an important role in congestion control and fairness guarantee among users. Edell et al. [21] implemented a usage pricing system and gave experiments illustrating that dynamic usage pricing can prevent congestion and improve the average network performance. Courcoubetis et al. [27] proposed intelligent agents to decide network usage, based on network conditions and users' payment willingness. This simplifies users' utility optimization process.\\
\indent After analyzing the  features of flat and usage pricing models, Altmann and Chu [23] proposed a hybrid pricing model that combines two. In this novel model, users enjoy basic services at a basic flat rate, while higher bandwidth demands will be charged by usage. The experimental data analysis indicates that such pricing model can improve network performance and increase ISP revenue. Obviously, such pricing concerning fixed and usage cost will benefit all the participants.\\
\indent Recently, with the continuous development of high-bandwidth required applications and P2P content distribution technologies, the overall users' bandwidth demands increase dramatically. Consequently, increasingly differentiated usage patterns make the fairness problem even more serious, which indicates charging heavy-load users according to usage is more reasonable [26]. However, in terms of P2P applications' providers who encourage users to participate in content sharing, such charging scheme will go contrary to their goals. So, more complicated interactions between P2P application providers and ISPs are to be carefully studied. In addition, other problems still need to be addressed, such as the privacy issues in processing audit and statistics [22] and the charging problem caused by users' non-expected traffic (such as ads and spams).
\subsection{Congestion pricing}
\begin{figure}
\centering
\includegraphics[width=0.5\textwidth]{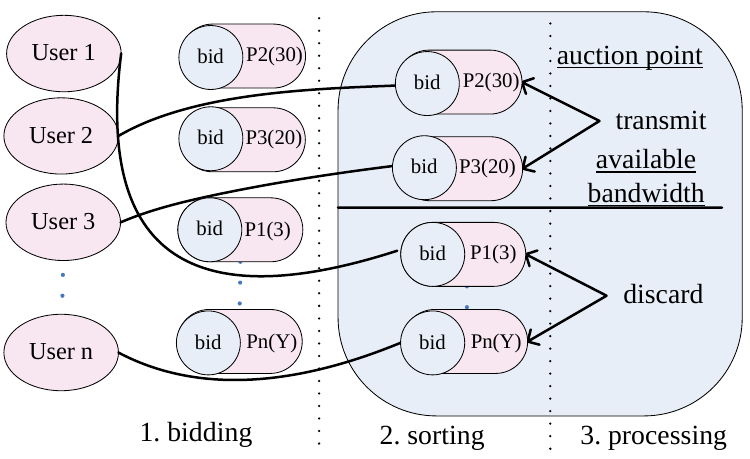}
\caption{ Smart market pricing.}\label{fig:5}
\end{figure}

The pricing models mentioned above cannot reflect individual traffic's impact on network, such as packet loss and delay. An intuitive understanding is that, too many concurrent network users will easily degrade network performance. For those who have accessed in network, the higher system load, the higher possibility of congestion. This also means that more external cost will be caused by users [18].\\
\indent Researchers expect pricing can constrain this negative external effect which is also called social cost. And the corresponding pricing is named congestion pricing [18]. Congestion pricing dynamically sets price that can reflect approximate real-time network resource usage and represent current social cost. Thus it can encourage users to adjust traffic demand which may avoid excessive resource usage. Therefore, congestion can be relieved or eliminated [29]-[37].\\
\indent However, measuring such social cost is not trivial. It cannot be directly calculated or measured as fixed or usage cost but need to detect users' perceived value of resources. In general network performance optimization articles, congestion cost is described by delay in M/M/1 queuing system [79]. In Mason and Varian's smart market [18] pricing mechanism, an auction based pricing method was proposed to measure and price such social cost. As shown in Fig.~\ref{fig:5}, the steps are as follows: (1) Users fill in bid fields for each packet on behalf of their willingness to pay for the packet transmission, e.g., P1(3) represents user 1's willingness to pay for its packet is 3; P2(30) means user 2 is willing to pay 30 for its packet; P3(20) means user 3 is willing to pay 20. (2) The routing node (auction point) receives the packets and sorts the packets according to the bid values. (3) Checking available bandwidth, the routing node sets marginal bid value as the market clearing price or threshold price, and decides which packets to be transmitted (or discarded). In this example, we see that if the node can only process two packets, the packets from user 2 and user 3 will be transmitted at the price of 20 for each packet, otherwise discarded.

\indent This can prevent congestion to some extent. Since the limited resources are allocated to people with high willingness to pay, the allocation will be more efficient. However, periodic bidding process and threshold price setting require additional technical supports from network protocols and hardware, making the method more technically complex. MacKie-Mason [35] further studied the advantages of smart market using generalized Vickrey auction mechanism [34] (i.e., when willingness to pay is personal privacy of an auction participate, the person with the highest bidding value will get the item at the second highest bidding value) to allocate scarce resources. The author concluded that the mechanism can promote truthful expression of users' utilities, and thus help network to attain service differentiation with different QoS levels. This kind of congestion pricing belongs to mechanism design (MD, [77]), which is always studied in incomplete information game theory area. We leave out more details here.\\
\indent There are also some pricing methods using congestion to set price levels (such as shadow pricing [30][31] and congestion discount [36]) and the relevant specific implementation mechanisms (such as congestion feedback based on TCP explicit congestion notification ECN). All aspects involved aim to implement efficient price-aware network resource usage which can shift the traffic from peak time to non-peak time, and thus reduce congestion possibility. In fact, time varying usage-based pricing can also achieve a certain level of congestion control [21], though it may not base on the analysis of social cost. Ykusel and Kalyanarama [37] analyzed the relationship between time granularity of congestion pricing and the resulting congestion level through experiments. They concluded that when the price interval is more than 40 times of RTT, the price can hardly affect congestion. So they suggested 2-3 seconds to be the appropriate pricing interval. However, such fine granularity of congestion pricing is not easy to implement in the real Internet.
\subsection{Discussion}
This section describes basic pricing models based on cost analysis in traditional \mbox{best-effort} network. They are gradually proposed and thoroughly studied along with the increase of network resource usage. Obviously, with the increasing importance of pricing in effective network resource management, pricing models will consider more factors and be more complex. From performance optimization perspective, this section describes pricing models with nearly different functions. In a flat pricing model, the fee is generally constant in a long period of time and is used to recover the fixed cost. Usage-based fee is charged to recover usage cost. It can be adjusted to reflect network congestion and thus plays a role in congestion control. Congestion pricing is proposed to measure and charge for congestion. It is a kind of dynamic pricing where price is dynamically adjusted to congestion.\\
\indent In fact, these three pricing models are not orthogonal, which means although they reflect different pricing factors, their functions can be overlapped to some extent. For example, ``two part tariff'' [19] was proposed as a combination of flat pricing and usage-based pricing. It can reduce congestion to some extent. In addition, congestion price mainly reflects the marginal cost of lacked resources. It can also be interpreted as the potential benefit increase of network users if there is one more resource unit. Therefore, congestion price is closely related to the timely network resources usage.

\section{Pricing mechanisms based on service type}
With more emphasis on QoS and network efficiency, services tend to be distinguished by data flow checking. This can help to achieve differentiated levels of QoS [62]. As a result, network service types can be divided into \mbox{best-effort} service and QoS mechanisms related services. Further, it is important that pricing models should be compatible with network service types [38]. This means that for different service types, pricing models should be suitable for charging. And there should have mechanisms to ensure the implementation of pricing. In this section, we describe pricing mechanisms to solve the above matching problem. And a brief analysis and evaluation will be given later.
\subsection{\mbox{Best-effort} service pricing}
In \mbox{best-effort} network, as ISPs generally do not implement additional QoS control mechanisms, there is nearly no QoS difference. Thus, ISPs adjust basic pricing models to affect resource usage while optimizing economic benefit. Pricing is always done at network edge, known as edge pricing [38][39]. It means that users' fees are calculated by the access network but not directly concerned with intermediate networks along the whole transmission path.\\
\indent Supporters hold the following beliefs. On the one hand, Internet users located in different autonomous system are often managed and charged by local ISPs. Thus it is more realistic to charge users at the access side. On the other hand, as a \mbox{best-effort} service, ISPs provide no QoS guarantee to whatever traffic traversed through their networks. So pricing at network edge is more reasonable [38][39].\\
\indent The basic pricing models suitable to edge pricing include flat pricing and usage pricing. For congestion pricing, because congestion could occur in any link along the transmission path, the price should be set according to path usage status. Thus it is not applicable to edge pricing. Moreover, for data packets, there may be multiple paths to select. But routing or path is not decided by users. So it is unfair to charge them for the path they use [38]. However, Shenker et al. [38] pointed out that edge pricing can still refer to approximate congestion and users' expected paths.\\
\indent Clark [39] further discussed localization method for non-local accounting and pricing, such as setting price for multicast users and pricing for receiver-paid applications. The method is based on resource reservation protocol (see Section 3.2.2), where the sender first chooses how much it will pay or what portion of cost to share with receivers. In [40][41], Clark suggested edge pricing could use estimated traffic instead of actual usage to charge users. And receivers can also state their willingness to pay. ISPs exchange traffic and revenue through agreements. Later, when bandwidth management devices [42] are added in the DiffServ architecture (see Section 3.2.3), the relatively dynamic edge pricing based on expectations or estimations is also being studied [43]. However, obviously, the edge pricing lacks influence on congestion control. Yuksel and Kalyanaraman [44] proposed a distributed dynamic pricing that is congestion sensitive and whose sensitivity and complexity are ranged between those of ``smart market'' and edge pricing.\\

\begin{figure}
\centering
\includegraphics[width=0.5\textwidth]{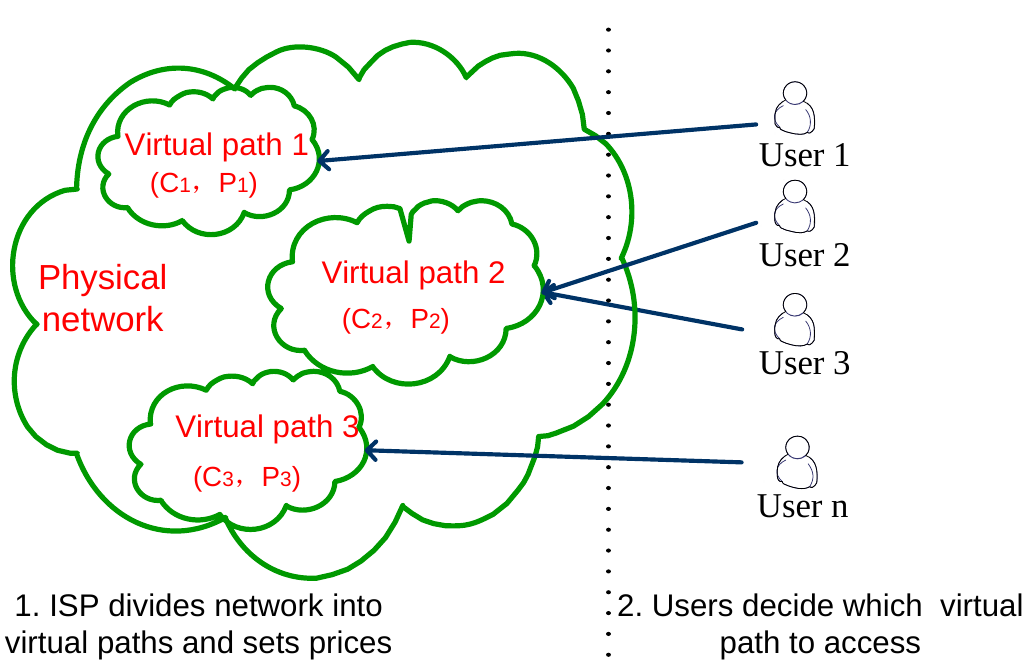}
\caption{ PMP pricing.}\label{fig:6}
\end{figure}

\indent Overall, edge pricing is applicable to \mbox{best-effort} network. ISPs can negotiate with users at access network based on expected congestion through predicting network states. Thus they arrive at pricing agreements. The pricing is easy to implement and can prompt flexible interaction between ISPs and users (such as ISPs can dynamically adjust price based on network conditions and users can adjust their QoS requirements according to their expected utilities). However, edge pricing is unable to conduct congestion control in the whole network due to networks' distributed characteristic. Although agreements exist, \mbox{network-wide} QoS guarantee or QoS differentiated services are hard to ensure.\\
\indent However, prioritized services can be implemented in \mbox{best-effort} network. Odlyzko proposed Paris Metro Pricing (PMP [55]) model, where a network is divided into several virtual transmission paths with different capacities and access prices. Thus users can expect to get differentiated services by accessing to different virtual paths. The main idea is that users can enjoy better performance at a higher possibility by paying more money. As shown in Fig.~\ref{fig:6}, the network is logically divided into channels or virtual paths with different transmission capacity $C$ and corresponding price $P$. In principle, selecting channels at higher payment will get better service as less competitors.

\indent The advantages of PMP are described as follows. As an edge pricing, paying for access based on expected performance is easy to implement. Since network providers divide users into different categories through charging, it is natural to achieve a certain degree of network resource management and differentiated services. The disadvantages are that the network will not maximize its usage efficiency and cannot ensure QoS. In addition, since it is very likely that different subnets use different pricing strategies, PMP applies only to a monopolistic network. So, if the model is to be extended to a complex network environment with many small networks, the price setting and revenue sharing should be consulted by those subnets. As to implementation, Odlyzko stated that users can simply choose different edge network providers according to different service qualities they provided. And for ISPs, within the network, routers are used to identify priority bits in packets and conduct priority-based scheduling or packet processing.\\
\indent Similar to PMP, Dube et al. [54] proposed a service differentiation method based on queue management. For users, each chooses and joins a queue according to its price and length. And for network server, it implements a priority-based queue scheduling in order to achieve differentiated resource allocations. Unlike in PMP, users here can estimate network congestion through queue lengths, and choose a service queue based on estimated congestion and its price. It is a profit maximization dynamic pricing model. Dube et al. used Markov decision theory (MDP) to build up system model, and presented dynamic price adjustment algorithms.
\subsection{QoS guaranteed service pricing}
Facing unachieved QoS differentiation and corresponding low network resource usage efficiency, a lot of work has committed to study of differentiated services so as to enhance efficiency. Simple priority-based service and corresponding pricing were first introduced by Cocchi et al. [13][14], which revealed the relationship between QoS differentiation and resource usage efficiency. They proposed to add priority field in IP packet and achieve QoS through priority-based queuing and scheduling. The corresponding service pricing is thus being wildly studied [55]-[60].\\
\indent Then, with the progressive development of various network service types, to achieve QoS guarantee, various in-depth studies were conducted regarding network architecture based on resource reservation [15] and flow aggregation [16]. Also, related pricing models are studied and integrated into such \mbox{QoS-enabled} pricing mechanisms [45]-[51].
\subsubsection{Simple priority-based service pricing}
\begin{figure}
\centering
\includegraphics[width=0.3\textwidth]{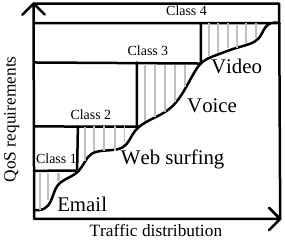}
\caption{ Service class division based on QoS requirements. [59]}\label{fig:7}
\end{figure}

To provide priority-based services, one reasonable way is to distinguish traffic by application's characteristics, as shown in Fig.~\ref{fig:7}. QoS based services can be divided into several classes. Generally, packets are set to different levels of transmission priority and help to achieve service distinction. The simplest way is using Type of Service (ToS) fields in IP packets to set priority levels. Such model is more realistic and implementable though QoS may not be guaranteed.\\
\indent With priority-based QoS differentiated services (similar to DiffServ in Section 3.2.3), a network can provide different service prices for each service class. And users can decide which service class to purchase. Since packet transmission for priority-based service depends on cooperation along the whole network path, a reasonable revenue sharing scheme may be required.\\
\indent For example, Cocchi et al. [13][14] believed that in a multi-class service coexisted network, if the resource is allocated based on applications' characteristics (or users' requirements), it will not only benefit users of all kinds of services, but also prompt an efficient network resource allocation. The basic idea is that for users to represent their utilities by filling priority fields in data packets. This will help network to implement user utility aware resource allocation (e.g., high priority packets will be processed earlier to avoid delay). Of course, packet transmission with higher priority will be charged at higher price.\\
\indent Specifically, here user utility is determined by price and QoS level $U=-V-C$, where transmission cost is represented by $C$, and $V$ measures the performance degradation (such as delay and packet loss rate). So applications such as FTP and Voice have different $V$ and thus will adopt different priorities. Therefore $p_{i,j}$ ($i=0,1$ and $j=0,1$) denotes four priority categories, where $i=1$ denotes using priority, and $j=1$ indicates the packet should not be discarded. Then if QoS is emphasized, the user will choose $p_{1,1}$ service class. And if the price is considered more, then $p_{0,0}$ service class will be more applicable. Obviously, for price levels, it will be $p_{0,0}<\{p_{0,1},p_{1,0}\}<p_{1,1}$. The corresponding relationships are: $\hbox{Email}\rightarrow p_{0,0}$, $\hbox{FTP}\rightarrow p_{0,1}$, $\hbox{Voice}\rightarrow p_{1,0}$ and the like. Simulation results show that differentiated service and pricing can incentivize users to choose appropriate service priorities. And the authors concluded that if revenue attained by such way is the same as what is gained without QoS differentiation, the former will achieve higher total utility. However, since service price is pre-set here, when idle resources exist, users will still pay more for prioritized services without QoS guarantee. So this is a preliminary work that uses ToS field to differentiate services and thus price differently.

\indent Similar to Cocchi et al, Donnell and Sethu [53] also suggested setting priorities or service classes for data packets by end user systems. Then, routers allocate them into different queues to ensure various service priorities. As to pricing implementation, the price field of a packet is filled in, which represents the payment for such transmission. Then when the packet reaches its destination, the price information is copied to ACK and returned to the sender. So the user (sender) can determine its sending rate and dynamically select the service class based on the received price information in ACK.\\
\indent Gupta et al. [56][57] proposed a more complex dynamic priority-based pricing mechanism, and designed a real-time external price calculation method based on the degree of congestion in multi-class service environment. Their simulation showed that dynamic pricing can significantly improve network performance and increase revenue. In order to avoid users to distribute traffics into non-matching service classes, [57] studied how to set appropriate price to encourage users in matching traffic type and service class in multi-class service network.\\
\indent Priority-based service pricing can achieve average performance differentiation if the price and traffic are relatively stable during a long time period. However, in the short term, it is likely that a high-priority service indeed experiences more packet loss, longer delay, serious congestion and so on. To solve this problem, [59][60] studied the proportional differentiated service model which provides a relatively dynamic bandwidth division scheme. The main idea is that, as an expansion of \mbox{best-effort} service type, the model will not strictly set bandwidth for each service class. Instead, it will use proportional performance guarantee to achieve predictable and controllable QoS distinction (based on well designed packet scheduling and packet discard mechanism). Compared with the fixed priority service, the corresponding proportional pricing model is more applicable to such service models.
\subsubsection{IntServ-based service pricing}
\begin{figure}
\centering
\includegraphics[width=0.5\textwidth]{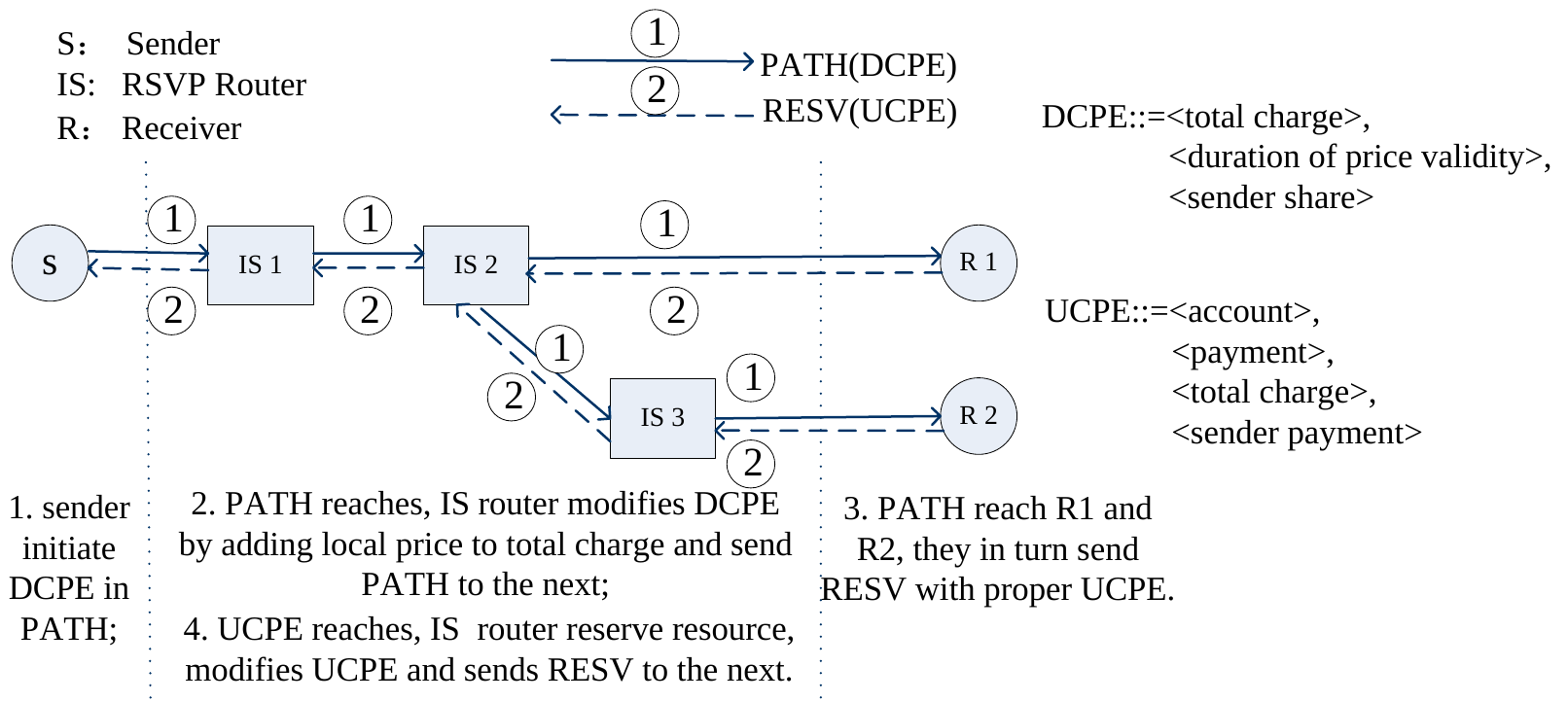}
\caption{Example of pricing session based on RSVP. [45]}\label{fig:8}
\end{figure}

In \mbox{best-effort} network and simple priority-based service network, QoS is not guaranteed. Accordingly, pricing usually depends on actual cost or resource usage. In contrast, this section will describe Integrated Service (IntServ [15]) mechanism, which achieves QoS guarantee from the perspective of resource reservation. Thus the corresponding pricing is extended from edge network to the entire resource reservation or QoS guaranteed path.\\
\indent IntServ bases on end-to-end Resource Reservation Protocol (RSVP [17]) to reserve resources for each flow. It is a single-flow based architecture that can provide end-to-end QoS guarantee. The overall mechanism needs all routers to process each flow's signaling messages, maintain its path and resource reservation status on control path, and perform flow-based classification and scheduling on data path. More specifically, based on packet transmission control, routers convert IP packets to traffic flows first. Then \mbox{RSVP-enabled} routers establish or dismantle resource reservation status of each flow according to their judgments on whether the path has sufficient resources to meet each incoming flow's QoS requirements. If met, based on packets' statuses, they implement QoS routing, corresponding scheduling and other controls to ensure the required QoS.\\
\indent Karsten et al. [45]  studied a pricing mechanism applicable to RSVP, as shown in Fig.~\ref{fig:8}. The main idea is to add price related information to regular RSVP messages and thus to achieve resource reservation and pricing conciliation. Specifically, the authors added Downstream Charging Policy Element (DCPE) in PATH message and Upstream Charging Policy Element (UCPE) in RESV message, where PATH and RESV are both regular RSVP messages (the description of DCPE and UCPE can be found in  Fig.~\ref{fig:8}). Then, the mechanism works as follows: first of all, it is sender $S$ that describes the flow's characteristics in PATH message and initiates DCPE to show its share of payment in the whole transmission (in $\left<\hbox{sender}\ \hbox{share}\right>$ field in DCPE). Then, each intermediate RSVP router (IS) who receives this information will modify DCPE by storing its local price into $\left<\hbox{total}\ \hbox{charge}\right>$, fill duration time information, and pass on PATH message. When PATH messages reaches $R1$ and $R2$, if any receiver accepts the service with such charging information, it sends RESV messages back with filled UCPEs. Specifically, the receivers calculate how much to pay based on the received DCPEs. They set $\left<\hbox{payment}\right>$ in UCPEs to show their cost sharing and copy the $\left<\hbox{total}\ \hbox{charge}\right>$ fields. When RESV reaches IS, IS reserves resources, modifies $\left<\hbox{sender}\ \hbox{payment}\right>$ information and passes on RESV. Upon sender $S$ receives RESV eventually, the $\left<\hbox{payment}\right>$ field carries the total charge paid by receivers. The $\left<\hbox{sender}\ \hbox{payment}\right>$ shows the fraction of charge on the sender, and $\left<\hbox{total}\ \hbox{charge}\right>$ carries the sum of all charges for this resource reservation. Obviously, this pricing mechanism has much flexibility in sharing cost between senders and receivers. And thus it can support pricing for many applications such as one or two side pay.

\indent Similarly, Clark [46] proposed a zone-based charging or cost sharing model. In this model, a willingness to pay information is inserted into an IP packet to show whether the two sides (sender and receiver) are willing to pay for high quality of services. However, it gave no more study on dynamic pricing and QoS class based pricing. Fankhauser et al. [47] proposed a RSVP-based accounting and charging protocol which is applicable to IntServ architecture. The authors showed such implementation can support local pricing models well using two pricing models. One is auction-based pricing model (adding bidding field in the RESV message), and the other is a congestion sensitive usage-based pricing model. However, it needs to assume that the network performs static routing which will not be affected by price, and each pricing node in the network prices at the same pace.\\
\indent In fact, flow-based resource reservation is hard to achieve. It needs to realize flow-based access control, QoS routing and related scheduling which will bring in huge system cost, and thus is very complex. Therefore, the realization of IntServ with QoS guarantee is not common, and only few applications exist. The improved IntServ and the corresponding pricing models are also under research.
\subsubsection{DiffServ-based service pricing}
\begin{figure}
\centering
\includegraphics[width=0.5\textwidth]{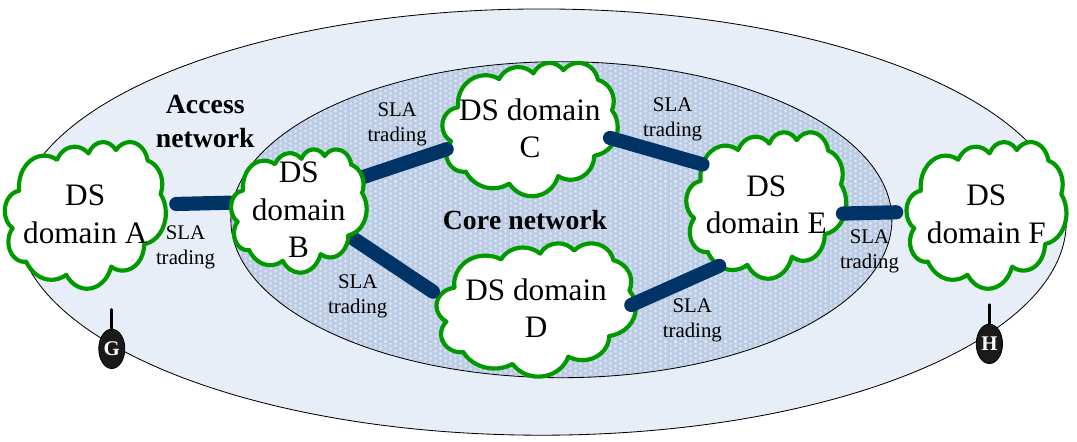}
\caption{Example of ISP networks at access and core levels. [48]}\label{fig:9}
\end{figure}

As \mbox{RSVP-based} IntServ architecture has high complexity and less scalability, Differentiated Services (DiffServ[16]) architecture is then proposed by IETF. Accordingly, the corresponding pricing is widely studied.\\
\indent In DiffServ architecture, complex flow control mechanism is realized at boundary nodes of the network. Thus service mechanism of network inward nodes is simplified. Specifically, the boundary nodes use users' flow profiles and resource reservation information to conduct flow-classification, shaping and aggregation, resulting in flows divided into different flow aggregations. And the aggregation information is stored in DS (Differentiated Service) field of IP packets called Differentiated Service Code Point (DSCP). Then the internal nodes schedule and forward IP packets in accordance with DSCP in \mbox{packet-headers} which represent the specified QoS requirements or service levels. DiffServ is a hierarchical service structure. Each DS region adopts SLA (Service Level Agreement [16] , i.e., a service contract between a customer and a service provider that specifies the service a customer should receive) and TCA (Traffic Conditioning Agreement) [16] to conduct coordination and thus to provide cross-regional services. SLA clearly describes the supported service level and the allowed traffic volume in each service level, and TCA is used in detailed QoS negotiation.\\
\indent Pricing is usually based on SLA in DiffServ architecture. Since SLA can be a static or dynamic contract used to describe the specified QoS level on data path, the corresponding pricing can also change with SLA's variation pace. In static SLA, regular consultations are needed. While in dynamic SLA, users need signaling protocol (e.g., RSVP) to help request service dynamically. And transformation is needed to match service requirements with DSCP value (no matter by user or edge router). Then, accordingly, the price for differentiated service depends on SLA and actual network resource usage. Fankhauser and Plattner [48] proposed an implementation profile to describe resource transactions in networks, which is based on bandwidth broker to act as an SLA trader or negotiator. The essence is that through negotiation between bandwidth brokers of each adjacent ISP, an ISP can provide its neighbors with its own network resources as well as the resources it purchased from other adjacent ISPs. Therefore the \mbox{Internet-wide} communication can be achieved. For example, in core network, as shown in Fig.~\ref{fig:9}, there are six DS domains: A, B, C, D, E and F. Each DS domain represents an ISP. Then B may offer service with destination E to network A, if it has bought service to destination E from network C or D. And in access network, as shown in Fig.~\ref{fig:9}, if user G (in network A) and network A arrive at an SLA that G will communicate with user H in network F, then an end-to-end service can be attained by building up bilateral agreements step-by-step in the form of SLAs between adjacent networks.

\indent The above work mainly discussed how to conduct inter-domain resource transaction based on DiffServ architecture with SLAs. But it did not mention pricing individual users based on DiffServ and the exact price. Semret et al. [49] established a double-layer DiffServ-based market model which considered users, bandwidth brokers and bandwidth sellers in the market. Each service class has its own bandwidth broker which belongs to bandwidth seller. The authors concluded that competitions among bandwidth brokers will lead physical bandwidths to an effectively division for various classes of services. Users adopt SLAs to negotiate services and prices with bandwidth brokers. And driven by dynamic market, bandwidth division among various service classes will finally be stable and efficient.\\
\indent Similarly, Wang and Schulzrinne proposed a framework named Resource Negotiation and Pricing (RNAP) [50]. They pointed out that pricing for reserved resource should be conducted differently on two levels. In an edge network, users and ISPs negotiate based on single flow. And in the core network, users' requests with the same service level and consultation interval are aggregated to process together. Finally, network resources are allocated based on single flow in the edge network. In [51], Wang and Schulzrinne built an optimization model to study pricing and the corresponding implementation which introduced access control to aid resource allocation. And they analyzed the resulting resource utilization in a differentiated service network. The authors concluded that \mbox{congestion-sensitive} pricing combined with \mbox{user-controllable} traffic rate not only can achieve congestion control to a large extent, but also can guarantee QoS requirements of different service classes. Since all routers participate in congestion pricing along transmission path, their work is more complex than edge pricing by Yukesl [44].\\
\indent In [52], the authors proposed a pricing mechanism that differs the core/edge network pricing. They claimed to charge users in access side with a Time of Day (TOD) price which can dynamically reflect congestion degree in core networks. For core networks (as shown in Fig.~\ref{fig:9}), dynamic pricing based on congestion for differentiated services is studied, where adaptable prices are published as signs of core network congestion status. The advantages are as follows: (1) Since access control can be conducted in user end system or edge network, it reduces network control information transmission and simplifies the core network processes; (2) On the other hand, as this pricing is based on DiffServ and concerns economic objectives and resource usage efficiency, it is easy to achieve a certain level of economic efficiency when providing QoS differentiated services. So, it is a flexible, scalable and efficient pricing mechanism in DiffServ architecture.
\subsection{Discussion}
Based on two types of network services considering QoS or not (\mbox{best-effort} service and QoS guaranteed service), we introduce two kinds of pricing mechanisms in this section. For the former, edge pricing is a relatively suitable implementation, PMP has also been proposed as a variation. And for the latter, we introduce the pricing mechanisms proposed mainly within the scope of pricing, which serve two service architectures named IntServ and DiffServ.\\
\indent For \mbox{best-effort} service, Shenker et al. believed that if edge pricing uses expected congestion information, it can also achieve a certain degree of congestion control. Also, one can distinguish access bandwidths to achieve some kind of prioritized services. But both of them cannot assure the usage efficiency of resources and guarantee QoS or network performance.\\
\indent For \mbox{QoS-based} service pricing, as QoS is differentiated by packet processing based on service classification or resource reservation, which often needs support from devices or networks on the entire transmission path, and thus is more complex than the former pricing mechanism. Especially for IntServ pricing, as QoS is guaranteed based on resource reservations where the service mechanism itself is complex, the corresponding ricing process can be even difficult with higher complexity. However, when it comes to service differentiation or DiffServ, from the perspective of efficiency, to a certain extent, we can conclude that it facilitates the efficient use of resources (high QoS requiring packets are prioritized processed) and ensures fairness among users (i.e., which service class or agreement is chosen by users), though there is no assured QoS guarantee. Indeed, combining IntServ (in edge network) with DiffServ (in core network) to provide differentiated services can have low complexity and improve efficiency with a certain degree of QoS guarantee.

\section{Pricing methods}
The above sections introduced pricing models that decide price structure/factors and the service types based pricing mechanisms that decide how to match price models and services. In this section, we will introduce pricing methods which determine appropriate price levels.\\
\indent In microeconomics, the price level depends on market environments or structures (such as monopolistic or competitive network [56]), which is calculated based on related pricing theories in the field of microeconomics. In network research area, besides considering on the market, resource pricing is also affected by network service mechanisms, and price is settled through modeling of utility optimization interactions of various entities.\\
\indent We will introduce two main network pricing methods here: (1) System optimization models mainly based on network utility maximization (NUM [30][31]) framework; (2) Strategic optimization models, i.e., when setting prices or making other decisions, consider strategic behaviors of the others.
\subsection{Pricing based on NUM}
From an economic point of view, efficient market means total social surplus or the sum of service providers' surplus and users' surplus is maximized [61], which equals to the difference maximization between the value of resources to users and the cost of providers. Under different market environments, different conclusions can be drawn. We mainly introduce system utility (social surplus) optimization oriented pricing method for a single network based on the optimization theory. The system is consisted of users with different utility functions and a network with resource constraints [29]. In fact, this research line has a tremendous influence on communication networks. It prompted an in-depth understanding of network architecture and a guided protocol design for more efficient network resource usages.\\
\indent Kelly [30] proposed the concept of Network Utility Maximization (NUM) which is the initial work of Internet system
optimization. In his work of pricing and resource allocation, the main object is to find the price that can make the
total resource demand and supply in equilibrium. According to market pricing theory in [56], if a system is in \mbox{demand-
supply} equilibrium, the system utility or social surplus will be maximized. NUM framework can be described by three
optimization problems. The system optimization is a radical problem which can be first modeled as:
$\hbox{maxmize}\ \underset{s}{\sum}U_s(x_s)$ (the service provider's cost is ignored), where $x_s$ denotes the traffic rate
and $U_s$ denotes the value or utility of the traffic to the corresponding user. The constraints are: (1) $Hy=x$, where
$H_{s\times r}$ denotes the \mbox{source-destination} pair $i\in\{1, 2, \cdots , s\}$ is served by path $j\in\{1, 2, \cdots , r\}$, and vector $y=\left\{y_1, y_2, \cdots , y_r\right\}^T$ denotes the resources distributed to all \mbox{source-destination} pairs on each feasible path. This constraint means the whole distributed resources equal to $x_s$ for any user; (2) $Ay\leq C$, where $A$ is a \mbox{0-1} matrix telling whether the distributed resource is on the link, and the constraint means the sum of all the distributed
resources will be no more than link capacity $C$; (3) $x,y\geq 0$. The above can be rewritten as the whole problem (A):
%\begin{equation}
\begin{eqnarray}
  \nonumber \hbox{SYSTEM}[U,H,A,C]:\\
  \begin{array}{ll}
    \hbox{maxmize} & \underset{s}{\sum}U_s(x_s) \\
    \hbox{subsect to} & Hy=x,Ay\leq C \\
    \hbox{over} & x,y\geq 0 \\
  \end{array}
\end{eqnarray}
%\end{equation}

\indent As user utility is unknown to the system, solving (A) is equal to solving two \mbox{sub-optimization} problems. One is on the user side, based on the per unit traffic rate price $\lambda_s$. An user optimizes its surplus $U_s(m_s/\lambda_s)-m_s$ by deciding how much to pay $m_s$ (the rate can be indirectly decided by $x_s=m_s/\lambda_s$), shown in the following problem (B):
\begin{eqnarray}
  \nonumber \hbox{USER}_{s}[U_s;\lambda_{s}]:\\
  \begin{array}{ll}
    \hbox{maxmize} & U_s(m_s/\lambda_s)-m_s \\
    \hbox{over} & m_s\geq 0 \\
  \end{array}
\end{eqnarray}

\indent The other sub-problem is on the network side. According to users' feedbacks, the network conducts optimization process and refers to some fairness standards to allocate network bandwidth to different flows. Namely, given $m=\left(m_1, m_2, ... ,m_s\right)$, it tries to distribute bandwidth by maximizing $\underset{s}{\sum}m_s\log x_s$, which indicates dividing bandwidth based on weighted proportional fairness. Then the corresponding network optimization problem (C) is:
\begin{eqnarray}
  \nonumber \hbox{NETWORK}[H,A,C; m]:\\
  \begin{array}{ll}
    \hbox{maxmize} & \underset{s}{\sum}m_s \log x_s  \\
    \hbox{subsect to} & Hy=x,Ay\leq C \\
    \hbox{over} & x,y\geq 0 \\
  \end{array}
\end{eqnarray}
where $H$, $A$ and $C$ denote the network status with the same meaning in Eq. (1). The author pointed out that if $\forall s, U_s(\cdot)$ is concave function, then from [30] we know that this convex optimization problem has a unique optimal solution $x^*=\left(x_1^*, x_2^*, ... , x_s^*\right)$. And author showed for $\lambda^*=\left(\lambda_1^*, \lambda_2^*, ... , \lambda_s^*\right)$, and $m^*=\left(m_1^*, m_2^*, ... , m_s^*\right)$, $m_s^*=\lambda_s^* x_s^*$ holds for every $s\in S$. Then the three optimization problems are all solved with consistent solutions. The vector $x^*$ is the unique optimal allocating rate and $\lambda^*$ is the current optimal resource price vector.\\
\indent System optimization problem (A) can also be decomposed into other types of sub-optimal problems, but the essence is not changed. So we skip it here. Kelly [31] further discussed the stability of the above mentioned rate allocation algorithm when the system is added in random disturbance and time delay. As to concrete solutions to the problem, since Kelly mainly modeled the elastic system, where users' utilities are all concave functions (reasonable when modeling traditional data services, such as file transfer, which is not very sensitive to delay), optimal solutions can be got based on the convex optimization theory. Similarly, some work [31][70][73] also use concave utility function to build models. Besides, the authors in [32] discussed a method that uses underlying buffer management to implement \mbox{end-to-end} proportional resource allocation, which can support Kelly's work.\\
\indent Unlike the centralized resource allocation method, Ozdaglar and Srikant [74] pointed out that if resources are distributively allocated like the above algorithm, then achieving system goals requires: (1) The end users should adjust their rates according to congestion feedbacks sent from the network (indicated by price); (2) The network routers should calculate the price which can reflect congestion status of each link starting from the router; (3) The network should be able to return the congestion information (price) to users. However, while the elastic flow rate can be adjusted according to network conditions (worked as TCP), in engineering, how to control rate based on the price is still not resolved. Practically, services with network feedback mechanisms are able to adopt such pricing method.\\
\indent In addition, users' controlling rates based on network feedbacks are not easy to implement as discussed in [74]. Based on the fair \mbox{end-to-end} congestion control mechanism proposed by Mo and Walrand [10], La and Anantharam [63] proposed a distributed algorithm where users can determine their rate adjustments according to their perceived network status. In their work, each user pays for queuing delay caused to others by its own packets. The authors proved the convergence of the algorithm, and showed it can solve this system optimization problem. It pointed out that packet loss rate can be used to formulate the optimization model as well.\\
\indent In this class of system (or users) utility maximization pricing work, rate allocation is based on user's willingness to pay (concave utility function). However, in fact, such willingness will vary with different types of applications. For example,
for video and voice applications, if transmission rate is less than a certain value, user's experience will decline sharply (as shown in Fig.~\ref{fig:9}). This indicates that $S$-type utility function should be used to model user's utility. And thus the convex optimization framework of NUM will no longer apply. The resulting system can be seen as a hybrid service system, as shown in Fig.~\ref{fig:9}, which includes different flows described by various types of utilities. Therefore, the pricing and resource allocation problem becomes a difficult non-convex optimization problem which should deal with competitive flows with different service characteristics [64][72].

\begin{figure}
\centering
\includegraphics[width=0.3\textwidth]{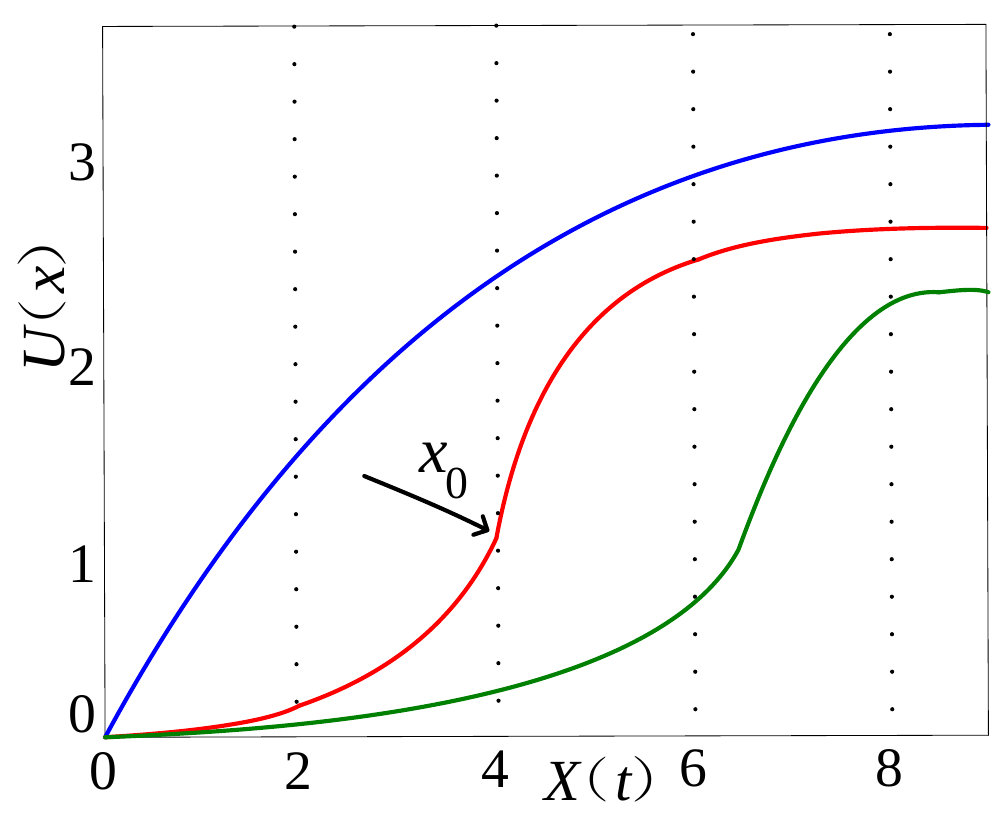}
\caption{Hybrid service system with various utility types.}\label{fig:10}
\end{figure}

\indent Jang-Won et al. [64][65] showed that in a real network environment (i.e., hybrid service systems), if the flows are all modeled by concave utility functions, then under the NUM framework, the resulting rate allocation will probably cause network congestion and high jitter. To achieve the optimal system resource usage when heterogeneous flows coexist, they studied distributed rate allocation and the corresponding pricing in a hybrid service system, and tried to design a reasonable incentive mechanism to incentivize users' transmission cancellations. Such user behavior is called ``self-regulate'' which is similar to the end system access control. The distributed algorithm is described as follows. For users, based on current price per unit rate, it decides the total transmission rate to maximize utility each time. And for network links, based on aggregated transmission rate, it solves network optimization problem and calculates unit rate price in next iteration. Mathematically, as the primal problem is non-convex, the duality gap may exist. This means the primal problem may not converge to the optimal, so the authors further designed asymptotical optimal resource allocation algorithm.\\
\indent Unlike the approximate optimal solution in [66], Chiang et al. [66] and Hande et al. [67][68] studied rate allocation optimization framework for inelastic flows, and presented the sufficient and necessary conditions for the convergence to the global optimum of the proposed distributed rate allocation algorithm. In contrast to the work by Jang-Won et al. [64], Chiang et al. [66] generalized the user utilities for different types of time-sensitive flows. They modeled them using non-convex optimization tools, and proposed heuristic access control algorithm and rate allocation algorithm. Similarly, considering the real-time flows, Hande et al. [67][68] introduced price-based distributed access control method and proposed a fair resource distribution method when various types of flows coexist. It emphasizes QoS-guarantee for the elastic flows and is realized by a proposed heuristic algorithm.\\
\indent In fact, since some researchers considered the access resource is most scarce and should be the study focus [70], in edge pricing model, the NUM framework has also been expanded and applied. For example, based on NUM, Hande et al. [70] studied the edge pricing in a monopoly market when ISP aims to maximize its revenue, and the user utility is modeled by standard $\alpha$-fairness based on different demand elasticity, namely:
\begin{equation}
u(x)=
\left\{
  \begin{array}{ll}
    (1-\alpha)^{-1}x^{1-\alpha}, & 0\leq\alpha<1 \\
    \log (x), & \alpha=1.
  \end{array}
\right.
\end{equation}
Unlike Kelly's work，the authors emphasized that in edge network, pricing structure can be a linear pricing combination composed by \mbox{time-related} flat fee and usage fee (e.g., $g+h\cdot x$). They analyzed each part's effect on ISP's revenue or what if using non-linear pricing.\\
\indent Currently, taking into account that the traffic is actually delivered from sender to receiver, the sender and receiver (supplier and demander) may have different utilities to such traffic. So, ISPs need to set a supply-demand balanced price to maximize its revenue. Hande et al. [69] extended the NUM framework by adding content providers (CPs) to the system optimization model. They concluded that no matter under which network marketing environment (complete competition or monopoly), considering the supply-demand relationship, if CPs are charged to compensate users, the overall system revenue and the utility of CPs will be sure to increase. The paper also discussed network neutrality (NN [94], that is, ISPs should not charge CPs by differentiating service quality or bases on content types) issue. If NN is equivalent to a constrained pricing for CP, then charging CPs can also improve system efficiency.
\subsection{Pricing based on game theory}
Within a single ISP network, system equilibrium based on \mbox{supply-demand} relationship can ensure optimal pricing which maximizes system utility in Section 4.1. This type of equilibrium is achieved through pricing where the ISP and users indirectly interact with each other. However, in real network, there are three types of relationships: ISP-ISP, ISP-users, and user-user. Most of them are modeled by considering their direct interactive effects.\\
\indent Game theory studies how individual decision is made considering others' actions. And it also predicts whether there exists an equilibrium under such strategic behaviors. The utilities in this type of model are directly affected by the other participants' strategies or preferences. Thus when studying the direct effects among network participates' behaviors, game theory is always used as a basic theory. Based on whether a binding agreement can be formed, games are divided into non-cooperative games [75][86] and cooperative games [83]-[85].
\subsubsection{Non-cooperative game model}
Considering non-cooperative games in network resource pricing and allocation, three levels of such interactions can be identified. (1) Competition among Multi-ISPs in network market. As users will purchase services from the most attractive ISP, so when an ISP decides pricing, it should consider the other ISPs' behaviors as well; (2) Leader-follower game between ISP and users. If ISPs consider users' reflection directly (e.g., not based on resulting demand as shown in Section 4.1, but beforehand consider how the price will affect the resulting demand), then this type of interaction can also be regarded as a game between ISPs and users; (3) Resource competition among users. Due to the externality caused by individual user to others, such internal impact can also be abstracted as a non-cooperative game. \\
\indent Multi-ISP interaction research has great challenges. Besides modelling similarities and difference among ISPs, impacts to underlay user behaviors should also be considered. Therefore, mature research results are still lacking today. In this section, we will mainly introduce the other two kinds of non-cooperative games (i.e., the above mentioned (2) and (3)). Two basic models frequently used here are $n$-person non-cooperative game model and leader-follower game model. The former emphasizes dynamic processes of a game, and the latter mainly considers static game equilibrium.\\
\indent It is reasonable to study above mentioned relationships in a single ISP network, since there indeed exists monopoly network market and thus the interference from other ISPs can be largely avoided. Then for modeling relationship (2), in a monopoly network market, single leader-follower game model (such as Stackelberg [77]-[81][89]) is always used. According to how much users' utility information known by the ISP, such work can be divided into two kinds: pricing with complete or incomplete information. For modeling relationship (3), $n$-person \mbox{non-cooperative} game is always used. Here each one's behavior affects the others' utilities, which is similar to externality mentioned in the foregoing discussion on congestion pricing.\\
\indent Generally, in the leader-follower network resource pricing model, a leader (ISP) sets price strategically, and the followers (users) act as price takers, who decide how much resource to buy mostly based on the given price. The point here is that when the leader decides price, it sets one that can maximize its revenue based on predicted users' reflections. In the $n$-person non-cooperative game, the stable state where none of participates wants to deviate from its behaviors when others' strategies are known, or NE (Nash Equilibrium [75]) is the major concern. An instance that combines the two models is presented by Basar and Srikant [78]-[81].\\
\indent Specifically, in [79], the authors used non-cooperative game models to study pricing issues in a \mbox{single-link} network. They built two layers of games: \mbox{non-cooperative} game related to resource competition among users and Stackelberg game where an ISP maximizes benefits within resource constraints based on predicting users' reflection. In the first layer model, each user maximizes its goal described by the following Eq. (5) to decide rate:
\begin{equation}
  F_i(x_i,x_{-i};p)=w_i\log(1+x_i)-\frac{1}{nc-\underset{j}{\sum}x_j}-px_i
\end{equation}
where $x_i$ is user's transmitting rate, $nc$ is link capacity, $w_i \log(1+x_i)$ is user's utility function,  $\frac{1}{nc-\underset{j}{\sum}x_j}$ represents congestion cost (i.e.,queuing delay computed using M/M/1 queuing model), and $p$ is the unit price charged by ISP.\\
\indent Using related theory, the authors prove that the users' non-cooperative game has NE, i.e., for any user $i$, the solution $x_i^*$ holds:
\begin{equation}
  \underset{0 \leq x \leq nc- x^*_{-i}}{\hbox{maxmize}}\ F_i(x_i,x^*_{-i};p)=F_i(x^*_i,x^*_{-i};p)
\end{equation}
It means that the decision made is the optimal one corresponding to all the others' optimal decisions.\\
\indent In the second layer game model, authors hypothesized that the ISP aims to maximize the benefits by solving Eq. (7), and thus obtain the unit resource price $p$.
\begin{equation}
  \underset{p\geq 0}{\hbox{maxmize}}\ L(p;\overline{x}^*(p)),L(p;\overline{x}):=p\cdot\overline{x}
\end{equation}
where $\overline{x}^*(p):=\sum_i x^*_j(p)$ represents the sum of all individuals' rates in NE of such \mbox{non-cooperative} game.\\ \indent The entire solving steps are as follows. Firstly, according to Eq. (5), it shows that adding up all utility functions of users will not change the NE point. The authors derived a user equivalent optimization problem in Eq. (8):
\begin{equation}
  F(x_1,\cdots,x_n;p)= \overset{n}{\underset{j=1}{\sum}} w_j\log(1+x_j)-\frac{1}{nc-\overline{x}}-p\overline{x}
\end{equation}
where all utilities are added together; Secondly, by solving the convex optimization problem, unique optimal solution $\overline{x}^*(p)$ can be obtained (notice that the solution is a function on price $p$). Thirdly, deduce the above solution to Eq. (7) as a single-variable optimization problem. And solve it directly can obtain the optimal price $p^*$. The authors then discussed what will happen for different link bandwidths here, and analyzed how the price, revenue and user's utility related with each other. They claimed that if the ISP expands bandwidth in proportion to the user number, then it will increase its revenue accordingly. However, it indicates that the users are expected to conduct congestion control based on the price and achieved service. And under certain circumstances, the solution will be consistent with Kelly's system optimal solution based on NUM model. The authors gave an extended discussion in the case of multi-link afterwards [78].\\
\indent Similar to the above mentioned \mbox{non-cooperative} game framework, Shen and Basar [81] extended the model to study \mbox{non-linear} optimal pricing in the cases of complete and incomplete information of users' utilities. They concluded that in the complete information (users' utilities are known by ISP) case, \mbox{non-linear} price can increase ISP's revenue by 38\% compared with revenue gained by linear price. While if users' utilities are unknown (incomplete information), there is about 25\% -40\% of the benefit loss. Li, Huang and Robert Li [71] also considered optimal pricing in a monopoly market with incomplete information. But they did not directly model users' non-cooperative behaviors.\\
\indent However, when an ISP prices users, in addition to considering the users' response strategies, the market environment is also taken into account. For example, in a \mbox{multi-ISP} market, ISPs compete for users, and their price are
affected by others. Thus the applicable game theory models will be very complex. Acemoglu and Ozdaglar [82] claimed that unlike in the monopoly case where system efficiency can be improved and the social optimal is achieved at the equilibrium, in the multi-ISPs competition game [61], the pure strategy NE may not exist (depending on cost function). And unlike the conclusions drawn from economics, the increasing competition will reduce system efficiency. Besides, the upper and lower bounds of possible loss are also discussed.
\subsubsection{Cooperative game model}
Historically, the well-studied cooperation game models in network resource pricing are the Nash Bargaining Game [83][84] and the Shapley value [85] model. These two models both belong to the axiomatic method, and thus their solutions satisfy certain properties. The former comes to Nash Bargaining Solution (NBS) with Pareto optimal property and a certain sense of fairness. The latter satisfies several good properties as well, which include well-formulated marginal contribution concept and the corresponding calculation methods. As a new trend, in recent years, such cooperative game models are studied and gradually applied to the modeling of network resource pricing.\\
\indent In [87], the authors assumed all network users have the same behavior characteristics and preferences. Therefore they simplified the problem as a game between a single user and one ISP. Based on theoretical analysis, they concluded that compared with the results in leader-follower game, Nash bargaining performed by ISPs and users can make the system operate at Pareto efficient (one cannot increase its utility without reducing others' utilities) operation point. In [88], based on Nash bargaining game, the authors studied network resource pricing and distributed allocation within a network with multiple heterogeneous users. We briefly introduce it as follows:\\
\indent First, the ISP faces a centralized resource allocation problem, in accordance with the concept of Nash bargaining. Such problem can be formulated as the following constrained convex optimization problem:
\begin{eqnarray}
  \begin{array}{ll}
  \hbox{maximize} &\overset{N}{\underset{i=1}{\prod}}(x_i-{MR}_i)\\
  \hbox{subsect to} & x_i\geq {MR}_i, \\
                      & x_i\leq {PR}_i,\\
                      & (Ax)_l\leq(C)_l.\\
  \end{array}
\end{eqnarray}
where $x_i$ is resource (rate) assigned to user $i$, ${MR}_i$ and ${PR}_i$ are the minimum and peak rate requirements of user $i$. Based on optimization theory, it is easy to know that there is a unique optimal solution. However, such central solution always brings in a lot of network communication burdens. Therefore, the authors proposed a distributed model where each user optimizes its utility with an added penalty $\alpha_ix_i$, and the aggregated rate is expected to ensure that the system can operate at the optimal point (Pareto optimal). Thus, for each user, it optimize Eq. (10) for rate selection:
\begin{eqnarray}
  \begin{array}{ll}
  \underset{x_i}{\hbox{maximize}} & \ln(x_i-{MR}_i)-\alpha_i x_i\\
   \hbox{subsect to} & x_i\geq {MR}_i, \\
                      & x_i\leq {PR}_i.\\
  \end{array}
\end{eqnarray}

\indent Similar to the leader-follower game in Section 4.2.1, the network here needs to solve the rate allocation problem which can maximize its revenue. Besides, the revenue is calculated by the sum of penalties, as shown in Eq. (11). The constraints conditions are the same as in Eq. (9).
\begin{equation}
  \hbox{maximize}\ \overset{N}{\underset{i=1}{\sum}}\alpha_i x_i\\
\end{equation}

\indent The authors designed and implemented an asynchronous distributed algorithm with the corresponding information exchange method, and showed that the solutions of Eq. (11) by network and Eq. (10) by users are equal to Nash bargaining solutions of the centralized problem in Eq. (9). The point is that such distributed method can maximize users' utilities as well as the network's revenue.\\
\indent Shapley value is mostly used in modeling for cost sharing or revenue distribution among multiple ISPs. Different from pricing directly based on usage information (such as pricing of core network [52] in Section 3.2.3), Shapley value emphasizes revenue distribution based on the contribution of each entity in a group. As an axiomatic method which ensures a unique solution, it has some special characteristics and is gradually applied to network resource pricing [92][93]. However, high calculation complexity is an obvious drawback (e.g., $N$ participants needs $2^N$ scale of computations). Besides, its requirements for a centralized allocation process also make it less scalable.
\subsection{Discussion}
We classified and summarized typical pricing methods of network resources based on two main research lines. The main points are as follows:\\
\indent (1) System optimization model mainly based on the NUM framework. Considering network traffic characteristics, it can be divided into: i) Optimization model for elastic flow system; ii) Optimization model for hybrid system where inelastic and elastic flows coexist.\\
\indent This class of work is usually based on supply-demand relationship. It aims to find the optimal price and rate allocation with balanced supply and demand where the maximal system efficiency is achieved. As inelastic traffics (such as real-time video and voice flows) emerge and largely increase, system optimization for hybrid network has drawn more and more attentions. Compared with elastic flow system optimization which has unique optimal solution shown by convex optimization theory, inelastic flows are always described by \mbox{$S$-type} utility function. Thus it turns the system problem into a complex \mbox{non-convex} optimization problem. Therefore, price-based access control is mainly introduced here to assist resource distribution. It generally includes two methods: users' self-regulation [65] based on their own utilities and the network conducted access control based on network efficiency [67]. Hande et al. [68] considered elastic flow protection in hybrid system. They believed that the elastic traffics are less competitive than the inelastic flows.\\
\indent In short, optimization-based modeling for hybrid system has high complexity, and especially hard to solve in real systems. Also, as on a network transmission path, access control policies of each link may be different, there lacks \mbox{well-designed} distributed \mbox{decision-making} mechanisms which can ensure system convergence to the global optimal rate allocation.\\
\indent (2) Strategic optimization model based on game theory. Based on two major branches of game theory: non-cooperative game
and cooperative game, we introduce some typical corresponding models used in network resource pricing.\\
\indent As modeling for strategic interactions, non-cooperative game model mainly discusses NE point and its characteristics. Cooperative game model we introduced here emphasizes the fairness criteria in sharing. The point is that the solution of the former may not be Pareto optimal, however, the latter sometimes needs constraints from a third party to ensure cooperation.\\
\indent Comparing the above system optimization model with the non-cooperative game model, it is obvious that different model ideas always need support from different theories. In NUM framework based system optimization, the equilibrium is achieved by indirect interactions between the network and users based on price. And in this process, ISP dynamically controls the system through pricing mechanism to help reach an optimal equilibrium. Non-cooperative game model directly analyzes pricing problem based on strategic behaviors of all participants, which can quickly determine whether the system has NE point or not. However, it is possible that the equilibrium point exists but is not achievable. Then, the uniqueness and stability of the existed NE will also be discussed mainly using optimization theory. In fact, if models have equal essential meanings for key parameters (e.g., revenue and cost), the results based on different basic theories (NUM or non-cooperative game) are nearly the same.

\section{Classification and comparison of Pricing Strategies}
In this section, based on pricing models, service mechanisms and price level setting methods, we conduct classification and comparison of the introduced typical pricing strategies shown in Table 1. (In order to describe the pricing for QoS guaranteed service, we add QoS contract in pricing model, which represents the achieved service and price agreements between ISPs and users).\\
\indent Early pricing models lack of theory basis. Most of them are based on experiments, and cannot cover a complete decomposition and classification. Some articles focused on studying pricing methods, but consider less about the underlying types of services. Since there are no separated pricing for QoS, we generally assume they are applicable for \mbox{best-effort} network, and make no special mark for them in Table 1. In addition, the QoS guaranteed types of services correspond to what we have described in Section 3.2. For pricing models, if both usage and access are chosen, it means that the pricing model is combined by the two.\\
\indent Considering implementation, pricing for different types of services inherently have different complexities. For \mbox{best-effort} network, pricing is always done at network edge, and needs less overhead cost. For QoS guaranteed services, since pricing relates with QoS along the whole serving path, it involves higher audition and accounting cost. But it can also achieve a certain degree of cost sharing (i.e., the sender and receiver consult on cost sharing). In short, the latter generally has a better QoS and higher network efficiency though at the cost of complexity.

\begin{table*}[htb]%\footnotesize
\centering
\caption{Classification of Pricing Strategies.} \label{table-operation}
\begin{minipage}[t]{0.9\textwidth}
\begin{center}
\begin{tabular}{|c|c|c|c||c|c|c|c||c|c|c||c|}
  \hline
    \multicolumn{4}{|c||}{\textbf{Pricing Model}} & \multicolumn{4}{c||}{\textbf{Service Type}} & \multicolumn{3}{c||}{\textbf{Pricing Method}} & \\
  \cline{1-11}
    Access & Usage & Con- & QoS & Best & \multicolumn{3}{c||}{QoS-guarantee} & System  & \multicolumn{2}{c||}{Game Model} & Example \\
  \cline{6-8}  \cline{10-11}
      &  & gestion & Contract & Effort & Priority & IntServ & DiffServ & Model & Non-co & Co & \\
  \hline
      & $\checkmark$ &  &  & $\checkmark$ &  &  &  &  &  &  &  [21][22]\\
   \hline
      & $\checkmark$ &  &  & $\checkmark$ &  &  &  & $\checkmark$ &  &  &  [25][27]\\
   \hline
      $\checkmark$ & $\checkmark$ &  &  & $\checkmark$ &  &  &  &  &  &  &  [23]\\
   \hline
      $\checkmark$ & $\checkmark$ &  &  & $\checkmark$ &  &  &  &  & $\checkmark$ &  &  [26]\\

   \hline
      $\checkmark$ & $\checkmark$ &  &  & $\checkmark$ &  &  &  & $\checkmark$ &  &  &  [19]\\
   \hline
      & $\checkmark$ & $\checkmark$ &  &  &  &  &  & $\checkmark$ &  &  &  [30][31]\\
   \hline
      &  & $\checkmark$ &  &  &  &  &  &  & $\checkmark$ (MD) &  &  [35]\\
   \hline
      &  & $\checkmark$ &  & $\checkmark$ &  &  &  & $\checkmark$ &  &  &  [36]\\
   \hline
      &  &  & $\checkmark$ & $\checkmark$ &  &  &  &  &  &  &  [38][39]\\
   \hline
      $\checkmark$ &  &  &  & $\checkmark$ &  &  &  & $\checkmark$ &  &  &  [55]\\
   \hline
      & $\checkmark$ & $\checkmark$ &  &  & $\checkmark$ &  &  & $\checkmark$ &  &  &  [54]\\
   \hline
      & $\checkmark$ &  &  &  & $\checkmark$ &  &  & $\checkmark$ &  &  &  [13][14]\\
      &              &  &  &  &              &  &  &              &  &  &  [53]\\
   \hline
      &  & $\checkmark$ &  &  & $\checkmark$ &  &  & $\checkmark$ &  &  &  [56]-[57]\\
   \hline
      &  &  & $\checkmark$ &  &  & $\checkmark$ &  &  &  &  &  [45][46]\\
   \hline
      & $\checkmark$ & $\checkmark$ &  &  &  & $\checkmark$ &  &  & $\checkmark$ (MD) &  &  [47]\\
   \hline
      & $\checkmark$ & $\checkmark$ &  &  &  &  & $\checkmark$ &  & $\checkmark$ (MD) &  &   [49]\\
   \hline	 						 	
      &  & $\checkmark$ & $\checkmark$ &  &  &  & $\checkmark$ & $\checkmark$ &  &  &  [51]\\
   \hline
      &  & $\checkmark$ &  &  &  & $\checkmark$ & $\checkmark$ & $\checkmark$ &  &  &  [52]\\
   \hline
      &  & $\checkmark$ &  & $\checkmark$ &  &  &  & $\checkmark$ &  &  &  [63]\\
   \hline
      & $\checkmark$ & $\checkmark$ &  & $\checkmark$ &  &  &  & $\checkmark$ &  &  &  [64]-[68]\\
      &              &              &  &              &  &  &  &              &  &  &  [72][73]\\
   \hline
      $\checkmark$ & $\checkmark$ &  &  & $\checkmark$ &  &  &  & $\checkmark$ &  &  & [70]\\
   \hline
      & $\checkmark$ &  &  & $\checkmark$ &  &  &  & $\checkmark$ &  &  &  69][71]\\
   \hline
      & $\checkmark$ &  &  & $\checkmark$ &  &  &  &  & $\checkmark$ &  &  [78]-[82]\\
   \hline
      & $\checkmark$ & $\checkmark$ &  & $\checkmark$ &  &  &  &  &  & $\checkmark$ &  [87][88]\\
   \hline
      & $\checkmark$ &  &  & $\checkmark$ &  &  &  &  &  & $\checkmark$ & [91][92]\\
   \hline
\end{tabular}\\[2pt]
\end{center}
\footnotesize In this table, the symbol $\checkmark$ in each row represents a feature hold by the pricing strategy example in the last column , and the symbol (MD) means mechanism design.
 \end{minipage}
\end{table*}

\section{Conclusion and future work}
In recent years, with the continuous development of high-bandwidth applications, content distribution technologies (such as CDN and P2P) are increasingly mature, and the network traffic surges. Thus network service quality has drawn more and more attentions. However, the engineering resource management and congestion control tend to have high technical difficulties, making network performance guarantee and maintenance even harder. Therefore, as a method that alleviates or resolves this problem by affecting active resource demand and usage, network resource pricing has important research values other than for ISPs to achieve economic goals. Besides, as QoS guaranteed services are getting more mature, and thus the pricing acts as an important auxiliary to incentivize technological progress, it is equally important to study pricing mechanism that is applicable to continuous renewal service types.\\
\indent In addition to pricing models and the corresponding service mechanisms, a complete pricing strategy also includes pricing methods deciding how much to charge. As shown in Fig.~\ref{fig:2}, we survey pricing issues from three different perspectives. We first introduce three basic pricing models: flat pricing, usage pricing and congestion pricing. And we conclude that with the development of network applications, research on pricing models turns more complex. Then, we introduce pricing mechanism which combines pricing model with service types. The mechanism aims to ensure pricing implementation under certain service types, such as transfer pricing information in DiffServ network. We notice that resource management for QoS differentiated networks with multi-class services mainly uses price-based access control. Then, from price level setting aspect, we highlight system optimization based on the NUM framework and strategic optimization based on game theory in a single ISP network. We conclude that the non-cooperative game models are often limited in related optimization theories to prove the existence of the Nash equilibrium. They are applicable only in part of (e.g., elastic flow system) models. And due to the incomplete information in such game, there is often a long distance from its actual application.\\
\indent To sum up, with the fast development of applications, service types, and corresponding theories, pricing related issues are constantly updated and studied. However, whichever pricing strategy we adopt, the basic pricing models and methods hardly change. For example, if the appropriate flat pricing brings in tolerable system efficiency loss, given its simplicity, such work should be revalued [89]. Through extensive study on network resource pricing strategies and deep analysis on the status quo, we can draw the following conclusions:

\begin{enumerate}
  \item Network resource or service pricing can be used as an effective tool to prompt technical progress, support QoS improvement, and/or enhance network efficiency economically.
  \item Economic oriented pricing strategy for network resource or service to price for QoS differentiation is still a hot research point, which also needs support from the corresponding complete service mechanisms.
  \item Pricing is expected to be scalable and easy to implement. It requires that besides mature theoretical models, well-designed mechanisms should also be implemented to help achieve pricing goals (such as maximizing resource usage efficiency or economic efficiency).
  \item As ISPs' revenue division will indirectly affect service quality and pricing of network users. Fair and implementable cooperation mechanism with \mbox{win-win} results among ISPs is also a hot topic for future research (e.g., in [91][92], fair revenue sharing models based on cooperative game theory were preliminary studied).
\end{enumerate}

\indent What's more, the models discussed above are unilateral market models whose network services include content provision. But if content providers and ordinary users (both have been modeled as users) are separately considered, then under such bilateral network market, pricing will involve more complex interactions. Also, the network neutrality concept [93] has been lately proposed, which causes more debates on whether the content should be charged differently. And we can infer that content-based pricing may also be discussed as part of pricing models in the near future.

\bibliographystyle{IEEEtran}

\end{document}